# Results of UBV Photoelectric Observations of the Early-Type Eclipsing Binary System XZ Cep

## M.I.Kumsiashvili, K.B.Chargeishvili, E.B.Janiashvili

E. K. Kharadze National Astrophysical Observatory, Georgia;
e-mail: kumsiashvili@genao.org, ketichargeishvili@yahoo.com, edik_var@yahoo.com

Results of the three-colour photoelectric observation of the close binary system XZ Cep, obtained at the Abastumani Astrophysical observatory, are presented.

XZ Cep (BD+66$^o$1512) is double-lines spectroscopic system consisting of late O and early B star. It was discovered by Schneller [1], who determined the type of variability and photometric elements. The history of photometric observations and their interpretations can be found in Glazunova and Karetnikov [2]. Giuricin et al [3] performed a light-curve analysis using the B and V photometry of Harvig [4]. Since the mass ratio was not determined at the time of their study, they examined several solutions for different mass ratios and concluded that the observations were best satisfied by the model in which the primary (brighter) component filled the Roche lobe. Mass can be transferred from the brighter component to the smaller detached secondary component, which is probably heavier.

A later photometric study of XZ Cep by Antokhina and Kumsiashvili [5] in which new UBV photoelectric observations were presented, used a spectroscopic mass ratio ($M_2/M_1$) of 0.78 Glazunova [6]. Modern methods of synthesis of the theoretical light curves of close binary systems were used.

The solution complies with the following model: the main the brighter but less massive component ($M_2$=14.2$M_\odot$) fills or nearly fills its own Roche lobe ($\mu_1$~1), temperature $t_2$=22 500 K. The second, more massive component ($M_1$=18.1$M_\odot$) is far from filling the Roche lobe ($\mu_2$~0.65) and has temperature $t_1$=31 600 K. In the main minimum, it is totally eclipsed by the brighter component. We note that there is no contradiction that the deeper minimum corresponds to the eclipse of the fainter star. Since the temperature of the companion is higher, there is a considerable affect of "reflection" from the side of the main component facing it. In the minimum of brightness the main component is towards us by its other cooler side, that is, the star $M_1$ and the hotter side of $M_2$ are eclipsed.

We note that in the phases 0.35-0.40, 0.6- 0.9 the theoretical light curves do not quite agree with the observations. This is obviously due to the influence of gaseous flows in the system, observed in these phases (Fig. 1).

These light-curves were analyzed again by Harries et al. [7] who also obtained a new radial-velocity curve of this system. Harries et al. confirmed that the system is a semidetached one and the cooler, less massive component is filling its critical Roche lobe.

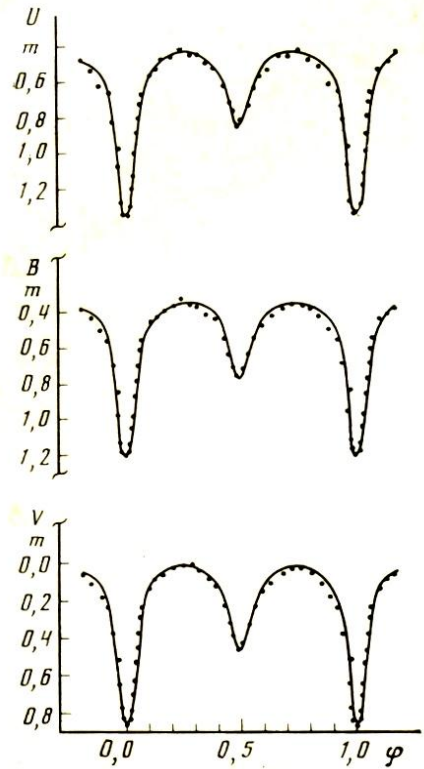

Fig. 1. Mean UBV light curves of XZ Cep (points) and theoretical light curves (lines).

On the bases above mention data they try to determine the location of the components of XZ Cep on the mass-log g diagram of fig.2, on which the evolutionary tracks (smooth curves) and isochrones (dashed curves) for individual stars from Schaller et al. [8] are also plotted. The locations of the primary (square) and secondary (triangle) components on this diagram show that XZ Cep, as are some other similar systems ($V_{448}$Cyg and $V_{382}$Cyg) is in a phase subsequent to a rapid transfer of mass.

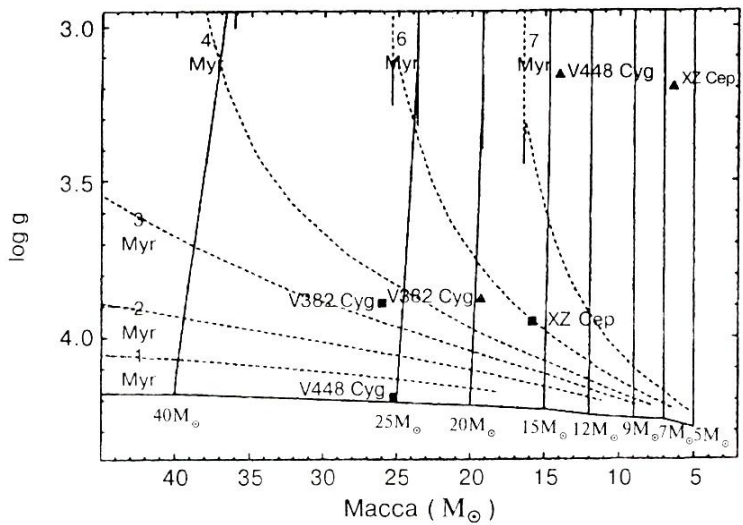

Fig. 2. Location of the components of XZ Cep on a mass-log g diagram.

Glazunova and Karetnikov [2] made their spectroscopic investigation on the bases of the spectra obtained at the 6 m. telescope of the Special Astrophysical Observatory Zelenchukskaya. They give spectral types for brighter component B1.5 II-III and for weaker component B1.1 III-V. These observations suggest the presence of envelopes in the system. The envelope round the secondary component is more sizeable and has a more complex morphology than the envelope of the primary component. The gaseous flow moves from the primary towards the secondary component.

The presence of a gaseous flow was detected in the interval of phases 0.7-0.9. The existence of circumstellar matter is also confirmed by the assumed change of intrinsic polarization of XZ Cep (Saute and Martel [9]).

Spectral and polarimetric study of close binary system XZ Cep was carry out also by Glazunova and Manilova [10]. Computation of the chemical composition of the atmosphere of primary component showed the existence of the hydrogen deficiency in it. The polarimetric measurements allowed establishing the sparse structure of the circumstellar envelope.

Recently, a probable variation in the polarization of eclipsing binary system XZ Cep was study by Kondon et al. [11]. XZ Cep shows large polarization degree up to 4.4%. The value had been almost constant throughout their observations at the whole wavelength. By them opinion the interstellar polarization must be dominant in the observed polarization of XZ Cep. In general, the polarization angle does not depend on wavelength for interstellar polarization. But in this case, the polarization angle is not constant. This slight change of polarization angle versus wavelength may be caused by the existence of two or more interstellar clouds with different properties, which lie between XZ Cep and us. They also mention that these variations do not depend on the orbital phase.

Follow above review the first thing to do was publishing of our individual photoelectric observations according to nights. This data was not publishing so far. The more so that duration of Abastumani observations and abundance of individual data attract one's attention.
The new three-color photoelectric observations were carrying out at the abastumani observatory on Mount Kanobili. These observations were made with 0.48 m reflector AZT-14A and AFM-6 photoelectric photometer. An FEU-79 electron multiplier as the detector of light and standard glass Schott filters for defining the UBV photoelectric system were used. The method of pulse counting was applied. The observations were made in 1972-1984. BD+65°1774 was used as the comparison and BD+66°1516 served as the reference star.

The observations made at Mount Kanobili in the UBV system cover the whole light curve and are so far the best and most extensive set of the observations of this star.

In each color separately 970 individual observations were carried out. The fact that in 1972,1973 observations of XZ Cep (6 nights altogether) were done with a self-recorder and beginning from 1975 a photon counter was used, is noteworthy. Accordingly, shifting of observations was observed in these years. In presented observations this fact was taken into account. Altogether 93 observational nights are at our disposal.
When drawing light variation curves the phases were calculated by elements Kreiner et al. [12] who performed a period study of the system, derived a new photometric ephemeris as follows

$$MinI = HJD2426033.421 + 5^d0972531E$$

The results of observations are given in the table and on the fig. 3. The moments of observations in Julius days, reduced to the solar centre, and phases are listed in columns one, two, three. Columns four, five and six shows the magnitude differences between the comparison and variable stars in yellow, blue and ultraviolet rays respectively.

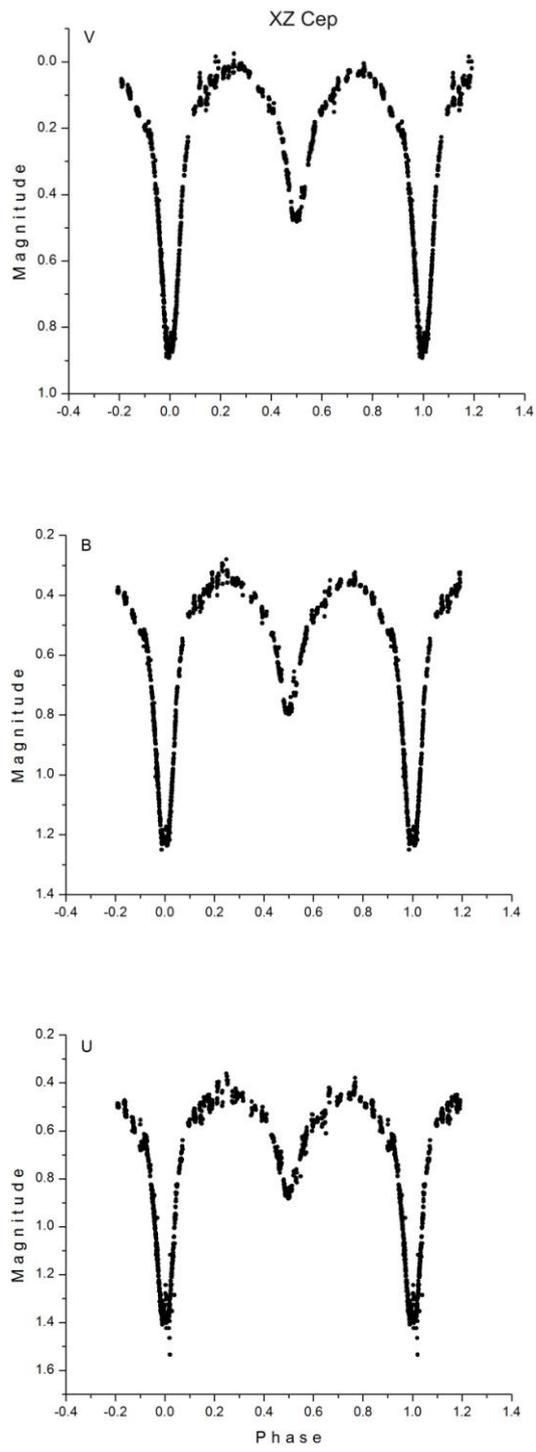

Fig. 3.

Table

| JD | phase | $\Delta V$ | $\Delta B$ | $\Delta U$ |
|---|---|---|---|---|
| 2441657.2304 | 0.1429 | 0.146 | 0.455 | 0.552 |
| 2441657.2387 | 0.1446 | 0.120 | 0.425 | 0.538 |
| 2441657.2470 | 0.1462 | 0.116 | 0.425 | 0.552 |
| 2441657.2540 | 0.1476 | 0.100 | 0.431 | 0.544 |
| 2441657.2596 | 0.1487 | 0.110 | 0.443 | 0.570 |
| 2441657.2637 | 0.1495 | 0.128 | 0.429 | 0.517 |
| 2441657.2832 | 0.1533 | 0.097 | 0.423 | 0.552 |
| 2441680.1684 | 0.6430 | 0.119 | 0.436 | 0.541 |
| 2441680.1725 | 0.6438 | 0.103 | 0.434 | 0.567 |
| 2441680.1795 | 0.6452 | 0.107 | 0.471 | 0.568 |
| 2441680.1836 | 0.6460 | 0.086 | 0.388 | 0.534 |
| 2441680.1899 | 0.6472 | 0.101 | 0.426 | 0.525 |
| 2441680.1948 | 0.6482 | 0.151 | 0.439 | 0.565 |
| 2441680.2016 | 0.6495 | 0.124 | 0.423 | 0.527 |
| 2441680.2059 | 0.6504 | 0.100 | 0.432 | 0.611 |
| 2441681.1600 | 0.8376 | 0.096 | 0.416 | 0.510 |
| 2441681.1642 | 0.8384 | 0.089 | 0.403 | 0.523 |
| 2441681.1684 | 0.8392 | 0.085 | 0.429 | 0.537 |
| 2441681.1718 | 0.8399 | 0.088 | 0.421 | 0.522 |
| 2441681.1767 | 0.8408 | 0.103 | 0.424 | 0.519 |
| 2441681.1823 | 0.8419 | 0.110 | 0.420 | 0.526 |
| 2441681.1878 | 0.8430 | 0.105 | 0.419 | 0.533 |
| 2441681.1920 | 0.8438 | 0.098 | 0.428 | 0.536 |
| 2441681.1954 | 0.8445 | 0.103 | 0.412 | 0.543 |
| 2441681.2003 | 0.8455 | 0.092 | 0.407 | 0.516 |
| 2442007.2362 | 0.8085 | 0.052 | 0.383 | 0.502 |
| 2442007.2411 | 0.8095 | 0.061 | 0.385 | 0.492 |

```
2442007.2466 0.8106  0.075  0.374  0.491
2442007.2515 0.8115  0.065  0.393  0.499
2442007.2585 0.8129  0.058  0.386  0.498
2442014.2145 0.1776  0.078  0.384  0.500
2442014.2201 0.1786  0.075  0.403  0.510
2442014.2263 0.1799  0.000  0.386  0.462
2442014.2319 0.1810 -0.016  0.365  0.492
2442014.2374 0.1820  0.076  0.364  0.474
2442014.2423 0.1830  0.062  0.374  0.490
2442014.2478 0.1841  0.056  0.370  0.471
2442273.3474 0.0153  0.821  1.207  1.424
2442273.3572 0.0172  0.781  1.156  1.375
2442273.3641 0.0186  0.799  1.159  1.306
2442273.3697 0.0197  0.821  1.217  1.395
2442273.3752 0.0208  0.835  1.194  1.346
2442273.3801 0.0217  0.772  1.139  1.292
2442273.3849 0.0227  0.759  1.123  1.247
2442273.3898 0.0236  0.768  1.085  1.248
2442273.3960 0.0248  0.758  1.086  1.281
2442273.4016 0.0259  0.740  1.123  1.245
2442273.4058 0.0268  0.758  1.087  1.242
2442642.4642 0.4302  0.185  0.531  0.619
2442642.4746 0.4322  0.197  0.525  0.617
2442642.4802 0.4333  0.202  0.534  0.619
2442642.4864 0.4345  0.212  0.528  0.630
2442642.4975 0.4367  0.208  0.538  0.623
2442642.5045 0.4381  0.212  0.527  0.627
2442642.5093 0.4390  0.215  0.541  0.647
2442642.5177 0.4406  0.213  0.543  0.631
2442642.5239 0.4419  0.233  0.551  0.644
2442642.5385 0.4447  0.242  0.573  0.669
2442661.2509 0.1158  0.073  0.428  0.527
2442661.2551 0.1166  0.079  0.425  0.519
2442661.2606 0.1177  0.047  0.409  0.525
2442661.2641 0.1184  0.034  0.410  0.518
2442661.2676 0.1191  0.065  0.415  0.522
2442661.2710 0.1197  0.058  0.402  0.507
2442661.2752 0.1206  0.069  0.423  0.510
2442661.2801 0.1215  0.061  0.422  0.522
2442663.3086 0.5195  0.408  0.688  0.787
2442663.3121 0.5202  0.438  0.726  0.771
2442663.3156 0.5209  0.380  0.676  0.769
2442663.3191 0.5216  0.388  0.655  0.774
2442663.3232 0.5224  0.393  0.735  0.768
2442716.2997 0.9155  0.184  0.515  0.614
2442716.3060 0.9167  0.181  0.515  0.637
2442716.3101 0.9175  0.211  0.531  0.629
2442716.3150 0.9185  0.206  0.523  0.643
2442716.3192 0.9193  0.206  0.525  0.646
2442716.3282 0.9211  0.219  0.537  0.646
2442716.3344 0.9223  0.245  0.537  0.655
2442716.3386 0.9231  0.215  0.532  0.658
2442716.3428 0.9240  0.225  0.553  0.662
2442716.3490 0.9252  0.224  0.548  0.668
2442716.3542 0.9262  0.217  0.556  0.665
```

```
2442716.3574 0.9268 0.222  0.557  0.669
2442749.1575 0.3617 0.086  0.393  0.499
2442749.1616 0.3625 0.083  0.407  0.508
2442749.1658 0.3633 0.094  0.404  0.506
2442749.1713 0.3644 0.086  0.406  0.496
2442749.1776 0.3656 0.084  0.408  0.505
2442749.1825 0.3666 0.091  0.400  0.507
2442749.1866 0.3674 0.104  0.406  0.512
2442750.1456 0.5555 0.269  0.580  0.687
2442750.1497 0.5563 0.260  0.586  0.678
2442750.1539 0.5572 0.264  0.578  0.673
2442750.1581 0.5580 0.261  0.574  0.671
2442750.1636 0.5591 0.264  0.586  0.676
2442750.1671 0.5597 0.255  0.572  0.672
2442750.1706 0.5604 0.253  0.566  0.669
2442750.1761 0.5615 0.245  0.566  0.667
2442750.1803 0.5623 0.247  0.560  0.661
2442750.1837 0.5630 0.237  0.559  0.664
2442750.1872 0.5637 0.244  0.562  0.657
2442750.1907 0.5644 0.240  0.551  0.647
2442750.1949 0.5652 0.229  0.544  0.653
2442750.1969 0.5656 0.238  0.561  0.637
2442750.2004 0.5663 0.223  0.548  0.648
2442750.2039 0.5670 0.219  0.549  0.647
2442750.2074 0.5677 0.218    -    0.660
2442752.1483 0.9484 0.393  0.729  0.880
2442752.1525 0.9493 0.396  0.745  0.881
2442752.1560 0.9499 0.414  0.739  0.855
2442752.1601 0.9507 0.427  0.743  0.865
2442752.1643 0.9516 0.433  0.762  0.872
2442752.1678 0.9523 0.434  0.764  0.895
2442752.1719 0.9531 0.443  0.776  0.909
2442752.1761 0.9539 0.448  0.782  0.916
2442752.1796 0.9546 0.461  0.789  0.942
2442752.1858 0.9558 0.470  0.811  0.944
2442752.1900 0.9566 0.492  0.820  0.945
2442752.1942 0.9574 0.487  0.824  0.959
2442752.1976 0.9581 0.496  0.827  0.975
2442752.2011 0.9588 0.517  0.847  0.990
2442752.2074 0.9600 0.515  0.845  1.006
2442752.2115 0.9608 0.516  0.864  0.997
2442752.2171 0.9619 0.542  0.885  1.008
2442752.2233 0.9631 0.546  1.006  1.032
2442752.2275 0.9640 0.570  0.896  1.044
2442752.2317 0.9648 0.566  0.922  1.047
2442752.2351 0.9655 0.585  0.918  1.052
2442752.2386 0.9661 0.587  0.939  1.074
2442752.2414 0.9667   -    0.941  1.092
2442752.2456 0.9675 0.607  0.962  1.076
2442752.2490 0.9682 0.618  0.962  1.095
2442752.2588 0.9701 0.638  0.975  1.139
2442752.2629 0.9709 0.650  0.999  1.151
2442986.4431 0.9133 0.212  0.543  0.657
2442986.4465 0.9140 0.217  0.537  0.654
2442986.4507 0.9148 0.223  0.571  0.675
```

```
2442986.4570  0.9161  0.224  0.549  0.669
2442986.4597  0.9166  0.219  0.523  0.656
2442986.4646  0.9176  0.220  0.541  0.654
2442986.4681  0.9182  0.215  0.549  0.669
2443096.2079  0.4475  0.233  0.579  0.685
2443096.2117  0.4482  0.240  0.585  0.686
2443096.2155  0.4489  0.273  0.587  0.676
2443096.2190  0.4496  0.281  0.591  0.685
2443096.2231  0.4504  0.274  0.594  0.691
2443096.2266  0.4511  0.283  0.593  0.699
2443096.2301  0.4518  0.285  0.582  0.703
2443096.2336  0.4525  0.290  0.592  0.709
2443096.2370  0.4532  0.293  0.596  0.704
2443097.1961  0.6413  0.101  0.427  0.521
2443097.2009  0.6423  0.098  0.425  0.524
2443097.2054  0.6431  0.103  0.425  0.518
2443097.2086  0.6438  0.102  0.423  0.529
2443097.2134  0.6447  0.108  0.429  0.531
2443097.2176  0.6455  0.094  0.422  0.526
2443097.2217  0.6463  0.105  0.429  0.530
2443097.2252  0.6470  0.099  0.416  0.522
2443097.2287  0.6477  0.085  0.423  0.515
2443097.2370  0.6493  0.082  0.425  0.529
2443098.2037  0.8390  0.092  0.408  0.525
2443098.2072  0.8397  0.094  0.405  0.509
2443098.2106  0.8403  0.093  0.420  0.522
2443098.2148  0.8412  0.089  0.424  0.526
2443098.2183  0.8419  0.098  0.421  0.523
2443098.2217  0.8425  0.096  0.414  0.530
2443098.2252  0.8432  0.106  0.427  0.542
2443098.2301  0.8442  0.111  0.423  0.533
2443130.1979  0.1157  0.124  0.459  0.551
2443130.2021  0.1166  0.119  0.446  0.535
2443130.2056  0.1173  0.130  0.447  0.553
2443130.2090  0.1179  0.126  0.448  0.544
2443130.2202  0.1201  0.108  0.439  0.539
2443130.2236  0.1208  0.112  0.432  0.547
2443130.2278  0.1216  0.114  0.438  0.525
2443130.2313  0.1223  0.111  0.435  0.523
2443130.2347  0.1230  0.119  0.445  0.543
2443130.2382  0.1237  0.114  0.441  0.530
2443131.1820  0.3088  0.028  0.373  0.457
2443131.1854  0.3095  0.030  0.361  0.447
2443131.1882  0.3100  0.037  0.358  0.456
2443131.1939  0.3111  0.031  0.359  0.444
2443131.1972  0.3118  0.033  0.364  0.456
2443131.2007  0.3125  0.047  0.366  0.457
2443131.2090  0.3141  0.033  0.364  0.450
2443131.2125  0.3148  0.030  0.400  0.441
2443131.2160  0.3155  0.038  0.364  0.462
2443131.2188  0.3160  0.035  0.363  0.475
2443132.1770  0.5040  0.466  0.775  0.836
2443132.1805  0.5047  0.459  0.773  0.850
2443132.1839  0.5054  0.456  0.783  0.845
2443132.1874  0.5061  0.453  0.778  0.852
```

```
2443132.1916 0.5069  0.466  0.763  0.848
2443132.1992 0.5084  0.472  0.740  0.848
2443132.2020 0.5089  0.467  0.771  0.852
2443132.2062 0.5097  0.466  0.764  0.839
2443132.2089 0.5103  0.455  0.788  0.849
2443132.2124 0.5110  0.464  0.788  0.852
2443132.2159 0.5116  0.477  0.769  0.847
2443132.2207 0.5126  0.455  0.778  0.829
2443132.2235 0.5131  0.447  0.779  0.844
2443342.3819 0.7429  0.036  0.359  0.448
2443342.3854 0.7436  0.042  0.366  0.465
2443342.3889 0.7442  0.037  0.363  0.448
2443342.3931 0.7451  0.040  0.358  0.442
2443342.3958 0.7456  0.040  0.355  0.459
2443342.4007 0.7466  0.032  0.356  0.451
2443342.4049 0.7474  0.029  0.351  0.453
2443342.4104 0.7485  0.033  0.355  0.455
2443342.4146 0.7493  0.038  0.351  0.468
2443342.4181 0.7500  0.033  0.359  0.452
2443346.3861 0.5284  0.409  0.734  0.773
2443346.3889 0.5290  0.398  0.726  0.809
2443346.3931 0.5298  0.386  0.692  0.797
2443346.3965 0.5305  0.402  0.700  0.801
2443346.4014 0.5314  0.396    -    0.761
2443346.4049 0.5321  0.386  0.696  0.775
2443346.4083 0.5328  0.378  0.696  0.816
2443400.3720 0.1196  0.127  0.443  0.540
2443400.3755 0.1203  0.125  0.448  0.536
2443400.3790 0.1210  0.125  0.445  0.555
2443400.3825 0.1217  0.123  0.440  0.551
2443400.3873 0.1226  0.130  0.448  0.563
2443400.3908 0.1233  0.124  0.440  0.538
2443400.3943 0.1240  0.129  0.444  0.553
2443405.3402 0.0943  0.144  0.462  0.566
2443405.3437 0.0950  0.158  0.469  0.574
2443405.3478 0.0958  0.143  0.470  0.576
2443405.3555 0.0973  0.146  0.464  0.577
2443405.3589 0.0980  0.152  0.472  0.580
2443405.3624 0.0986  0.151  0.462  0.569
2443405.3673 0.0996  0.136  0.470  0.575
2443405.3708 0.1003  0.138  0.462  0.554
2443405.3749 0.1011  0.135  0.456  0.553
2443406.3423 0.2909  0.017  0.343  0.449
2443406.3464 0.2917  0.016  0.351  0.468
2443406.3506 0.2925  0.014  0.346  0.452
2443406.3541 0.2932  0.015  0.344  0.440
2443406.3583 0.2940  0.021  0.352  0.462
2443406.3610 0.2946  0.016  0.349  0.449
2443406.3645 0.2952  0.019  0.347  0.448
2443406.3680 0.2959  0.023  0.350  0.458
2443406.3714 0.2966  0.024  0.363  0.445
2443406.3742 0.2971  0.027  0.363  0.452
2443407.3924 0.4969  0.476  0.788  0.876
2443407.3959 0.4976  0.465  0.780  0.861
2443407.3993 0.4983  0.466  0.789  0.883
```

```
2443407.4035 0.4991 0.470 0.791 0.873
2443407.4077 0.4999 0.459 0.796 0.871
2443407.4104 0.5004 0.470 0.797 0.870
2443407.4139 0.5011 0.481 0.775 0.864
2443407.4174 0.5018 0.477 0.785 0.852
2443407.4215 0.5026 0.480 0.790 0.862
2443407.4250 0.5033   -   0.790 0.871
2443408.4521 0.7048 0.027 0.358 0.467
2443408.4570 0.7058 0.035 0.366 0.476
2443408.4611 0.7066 0.028 0.355 0.468
2443408.4653 0.7074 0.029 0.369 0.473
2443408.4688 0.7081 0.025 0.356 0.474
2443408.4736 0.7090 0.025 0.359 0.472
2443408.4771 0.7097 0.026 0.347 0.458
2443408.4806 0.7104 0.030 0.358 0.460
2443408.4840 0.7111 0.031 0.358 0.448
2443408.4875 0.7117 0.030 0.350 0.449
2443409.4431 0.8992 0.197 0.522 0.664
2443409.4479 0.9002 0.194 0.518 0.673
2443409.4514 0.9008 0.198 0.527 0.574
2443409.4549 0.9015 0.199 0.523 0.555
2443409.4591 0.9024 0.205 0.523 0.682
2443409.4625 0.9030 0.201 0.529 0.675
2443409.4660 0.9037 0.190 0.522 0.661
2443409.4695 0.9044 0.194 0.532 0.646
2443409.4736 0.9052 0.208 0.525 0.678
2443409.4771 0.9059 0.208 0.527 0.679
2443420.2377 0.0169 0.815 1.182 1.365
2443420.2419 0.0178 0.805 1.169 1.340
2443420.2454 0.0185 0.807 1.162 1.334
2443420.2488 0.0191 0.805 1.152 1.321
2443420.2537 0.0201 0.790 1.152 1.324
2443420.2586 0.0210 0.777 1.140 1.302
2443420.2620 0.0217 0.759 1.128 1.282
2443420.2655 0.0224 0.764 1.121 1.286
2443420.2690 0.0231 0.756 1.109 1.271
2443420.2724 0.0238 0.751 1.098 1.266
2443423.2732 0.6125 0.135 0.454 0.558
2443423.2767 0.6131 0.140 0.462   -
2443423.2802 0.6138 0.135 0.448 0.559
2443423.2830 0.6144 0.122 0.456 0.563
2443423.2864 0.6151 0.135 0.460 0.564
2443423.2892 0.6156 0.132 0.455 0.556
2443423.2927 0.6163 0.136 0.448 0.544
2443423.2955 0.6168 0.130 0.442 0.557
2443423.2982 0.6174 0.129 0.447 0.559
2443423.3017 0.6181 0.132 0.448 0.551
2443424.2718 0.8084 0.057 0.388 0.486
2443424.2760 0.8092 0.063 0.391 0.495
2443424.2795 0.8099 0.061 0.388 0.485
2443424.2830 0.8106 0.059 0.390 0.502
2443424.2864 0.8112 0.056 0.387 0.489
2443424.2892 0.8118 0.066 0.389 0.496
2443424.2927 0.8125 0.060 0.382 0.495
2443424.2955 0.8130 0.063 0.384 0.503
```

```
2443424.2989 0.8137 0.064  0.390 0.489
2443424.3024 0.8144 0.068  0.388 0.502
2443425.2732 0.0048 0.853  1.219 1.378
2443425.2767 0.0055 0.858  1.209 1.394
2443425.2795 0.0061 0.848  1.214 1.402
2443425.2823 0.0066 0.845  1.219 1.380
2443425.2857 0.0073 0.847  1.211 1.392
2443425.2885 0.0078 0.845  1.217 1.378
2443425.2913 0.0084 0.848  1.226 1.375
2443425.2941 0.0089 0.840  1.224 1.380
2443425.2975 0.0096 0.845  1.229 1.391
2443425.3010 0.0103 0.841  1.211 1.382
2443425.3052 0.0111 0.842  1.218 1.376
2443425.3093 0.0119 0.843  1.212 1.374
2443425.3121 0.0125 0.837  1.214 1.379
2443425.3156 0.0131 0.836  1.213 1.376
2443425.3184 0.0137   -    1.199 1.378
2443428.2913 0.5969 0.152  0.468 0.581
2443428.2941 0.5975 0.160  0.469 0.578
2443428.2968 0.5980 0.154  0.486 0.578
2443428.3003 0.5987 0.155  0.468 0.584
2443428.3038 0.5994 0.151  0.492 0.572
2443428.3066 0.5999 0.156  0.473 0.573
2443428.3100 0.6006 0.153  0.490 0.570
2443428.3114 0.6009 0.143  0.477 0.555
2443428.3156 0.6017 0.152  0.477 0.570
2443428.3191 0.6024 0.135  0.471 0.556
2443429.2601 0.7870 0.041  0.367 0.470
2443429.2628 0.7875 0.037  0.369 0.467
2443429.2656 0.7881 0.036  0.367 0.474
2443429.2684 0.7886 0.040  0.369 0.470
2443429.2725 0.7894 0.042  0.367 0.477
2443429.2753 0.7900 0.036  0.370 0.473
2443429.2781 0.7905 0.038  0.365 0.463
2443429.2809 0.7911 0.045  0.365 0.459
2443429.2843 0.7917 0.042  0.368 0.476
2443429.2871 0.7923 0.040  0.369 0.463
2443430.2490 0.9810 0.776  1.141 1.307
2443430.2539 0.9820 0.779  1.148 1.334
2443430.2567 0.9825 0.802  1.156 1.330
2443430.2595 0.9831 0.802  1.168 1.345
2443430.2664 0.9844 0.815  1.187 1.347
2443430.2699 0.9851 0.813  1.188 1.351
2443430.2726 0.9856 0.825  1.189 1.362
2443430.2761 0.9863 0.834  1.194 1.355
2443430.2796 0.9870 0.841  1.203 1.368
2443430.2817 0.9874 0.844  1.204 1.372
2443430.2858 0.9882 0.844  1.212 1.384
2443430.2886 0.9888 0.841  1.221 1.391
2443430.2914 0.9893 0.847  1.220 1.409
2443430.2942 0.9899 0.852  1.227 1.401
2443430.2976 0.9905 0.857  1.220 1.383
2443430.3018 0.9914 0.852  1.224 1.399
2443430.3053 0.9920 0.856  1.222 1.405
2443431.3129 0.1897   -    0.325 0.490
```

```
2443431.3185 0.1908    -     0.331  0.508
2443431.3206 0.1912  -0.001  0.346  0.494
2443431.3226 0.1916   0.020  0.371  0.495
2443431.3261 0.1923    -     0.324  0.473
2443431.3282 0.1927    -     0.357  0.481
2443436.3581 0.1795   0.050  0.379  0.449
2443436.3643 0.1807   0.058  0.390  0.507
2443436.3692 0.1817   0.043  0.385  0.522
2443436.3726 0.1824   0.051  0.371  0.476
2443436.3754 0.1829   0.051  0.382  0.481
2443436.3789 0.1836   0.042  0.371  0.491
2443436.3824 0.1843   0.047  0.376  0.488
2443436.3858 0.1849   0.042  0.380  0.479
2443436.3893 0.1856   0.050  0.367  0.486
2443436.3942 0.1866   0.044  0.380  0.493
2443449.1804 0.6950   0.052  0.378  0.478
2443449.1872 0.6964   0.057  0.377  0.493
2443449.1907 0.6971   0.055  0.368  0.459
2443449.1942 0.6977   0.059  0.379  0.478
2443449.1976 0.6984   0.049  0.373  0.475
2443449.2018 0.6992   0.045  0.371  0.463
2443449.2053 0.6999   0.046  0.375  0.483
2443449.2081 0.7005   0.049  0.365  0.473
2443449.2115 0.7011   0.053  0.371  0.479
2443449.2143 0.7017   0.048  0.369  0.479
2443486.2482 0.9672   0.609  1.004  1.107
2443486.2517 0.9678   0.630  0.979  1.135
2443486.2552 0.9685   0.577  0.930  1.115
2443486.2582 0.9691   0.631  0.981  1.141
2443486.2635 0.9702   0.629  0.988  1.139
2443486.2670 0.9708   0.640  1.011  1.164
2443486.2705 0.9715   0.653  1.020  1.178
2443486.2732 0.9721   0.661  1.027  1.185
2443486.2774 0.9729   0.678  1.034  1.170
2443486.2819 0.9738   0.675  1.061  1.186
2443486.2836 0.9741   0.702  1.059  1.233
2443486.2906 0.9755   0.719  1.081  1.225
2443486.2948 0.9763   0.720  1.084  1.230
2443486.2982 0.9770   0.798  1.123  1.270
2443486.3010 0.9775   0.725  1.094  1.236
2443486.3045 0.9782   0.754  1.118  1.236
2443486.3080 0.9789   0.764  1.131  1.259
2443486.3114 0.9796   0.759  1.125  1.270
2443521.2490 0.8338   0.070  0.400  0.486
2443521.2553 0.8350   0.085  0.403  0.479
2443521.2587 0.8357   0.078  0.410  0.475
2443521.2629 0.8365   0.091  0.409  0.484
2443521.2671 0.8373   0.090  0.406  0.472
2443521.2705 0.8380   0.090  0.392  0.477
2443521.2733 0.8385   0.096  0.423  0.486
2443521.2775 0.8393   0.083  0.412  0.494
2443521.2810 0.8400   0.080  0.420  0.479
2443521.2844 0.8407   0.085  0.412  0.480
2443523.3635 0.2486   0.016  0.280  0.361
2443523.3711 0.2501   0.007  0.317  0.389
```

```
2443523.3739 0.2506  0.033  0.332  0.372
2443523.3816 0.2521  0.031  0.358  0.393
2443523.3850 0.2528 -0.025  0.317  0.407
2443523.3878 0.2534 -0.004  0.330  0.405
2443523.3913 0.2540  0.014  0.326  0.389
2443523.3947 0.2547  0.009  0.337  0.401
2443672.3588 0.4791  0.422  0.753  0.807
2443672.3630 0.4799  0.450  0.750  0.804
2443672.3665 0.4806  0.441  0.754  0.813
2443672.3706 0.4814  0.444  0.755  0.824
2443672.3741 0.4821  0.449  0.764  0.837
2443672.3769 0.4826  0.457  0.761  0.821
2443672.3810 0.4835  0.467  0.774  0.848
2443672.3845 0.4841  0.469  0.773  0.834
2443672.3873 0.4847  0.454  0.775  0.861
2443672.3908 0.4854  0.476  0.775  0.864
2443672.3949 0.4862  0.465  0.784  0.839
2443672.3984 0.4869  0.471  0.775  0.858
2443672.4019 0.4876  0.463  0.777  0.872
2443672.4047 0.4881  0.466  0.792  0.856
2443672.4074 0.4886  0.464  0.793  0.856
2443695.4430 0.0078  0.868  1.235  1.382
2443695.4471 0.0087  0.862  1.237  1.391
2443695.4513 0.0095  0.863  1.230  1.390
2443695.4555 0.0103  0.874  1.234  1.391
2443695.4589 0.0110  0.858  1.235  1.387
2443695.4624 0.0117  0.857  1.224  1.357
2443695.4659 0.0123  0.854  1.216  1.385
2443695.4686 0.0129  0.851  1.222  1.374
2443695.4714 0.0134  0.849  1.227  1.361
2443695.4756 0.0142  0.843  1.203  1.379
2443727.4234 0.2819  0.012  0.347  0.456
2443727.4269 0.2826  0.018  0.364  0.459
2443727.4311 0.2834  0.005  0.348    -
2443727.4345 0.2841    -    0.345  0.480
2443727.4422 0.2856  0.023  0.368  0.441
2443727.4449 0.2861  0.020  0.361  0.424
2443730.4063 0.8671  0.143  0.461  0.543
2443730.4104 0.8679  0.149  0.467  0.561
2443730.4139 0.8686  0.137  0.450  0.600
2443730.4174 0.8693  0.146  0.461  0.602
2443730.4208 0.8699  0.151  0.466  0.566
2443730.4243 0.8706  0.145  0.471  0.544
2443730.4278 0.8713  0.145  0.468  0.560
2443730.4313 0.8720  0.139  0.477  0.555
2443730.4347 0.8727  0.155  0.478  0.565
2443730.4382 0.8734  0.150  0.473  0.568
2443731.4182 0.0656  0.274  0.594  0.694
2443731.4216 0.0663  0.263  0.574  0.689
2443731.4251 0.0670  0.262  0.584  0.727
2443731.4293 0.0678  0.255  0.583  0.717
2443731.4327 0.0685  0.253  0.583  0.720
2443731.4362 0.0691  0.258  0.582  0.689
2443731.4397 0.0698  0.242  0.571  0.665
2443731.4432 0.0705  0.244  0.545  0.667
```

```
2443731.4466 0.0712 0.240 0.566 0.639
2443731.4501 0.0719 0.227 0.555 0.677
2443751.2856 0.9633 0.575 0.919 1.117
2443751.2898 0.9641 0.584 0.979 1.033
2443751.2933 0.9648 0.600 0.935 1.081
2443751.2960 0.9653 0.606 0.948 1.065
2443751.2995 0.9660 0.617 0.973 1.112
2443751.3030 0.9667 0.628 0.987 1.109
2443751.3072 0.9675 0.639 0.994 1.118
2443751.3106 0.9682 0.660 0.996 1.104
2443751.3134 0.9687 0.657 0.995 1.112
2443751.3169 0.9694 0.663 1.018 1.130
2443751.3204 0.9701 0.703 1.035 0.964
2443761.3699 0.9417 0.384   -     -
2443761.3740 0.9425 0.297 0.617 0.743
2443761.3768 0.9430 0.368 0.682 0.763
2443761.3810 0.9438 0.373 0.684 0.825
2443761.3851 0.9446 0.384 0.708 0.846
2443761.3879 0.9452 0.396 0.706 0.826
2443761.3914 0.9459 0.385 0.712 0.835
2443761.3956 0.9467 0.397 0.723 0.834
2443761.3990 0.9474 0.411 0.719 0.864
2443761.4025 0.9481 0.419 0.732 0.866
2443761.4067 0.9489 0.384 0.764 0.898
2443761.4129 0.9501 0.431 0.765 0.888
2443761.4157 0.9506 0.448 0.790 0.847
2443762.3928 0.1423 0.123 0.440 0.544
2443762.3969 0.1431 0.128 0.433 0.563
2443762.4004 0.1438 0.115 0.437 0.539
2443762.4039 0.1445 0.112 0.432 0.537
2443762.4081 0.1453 0.116 0.425 0.538
2443762.4108 0.1459 0.113 0.433 0.547
2443762.4148 0.1467 0.111 0.434 0.552
2443763.4442 0.3486 0.079 0.405 0.536
2443763.4476 0.3493 0.076 0.423   -
2443763.4518 0.3501 0.084 0.419   -
2443763.4546 0.3506 0.088 0.416 0.478
2443763.4581 0.3513 0.086 0.414 0.506
2443763.4608 0.3519 0.089 0.414 0.527
2443763.4643 0.3525 0.087 0.410 0.512
2443764.4484 0.5456 0.308 0.629 0.720
2443764.4526 0.5464 0.311 0.626 0.714
2443764.4554 0.5470 0.294 0.610 0.702
2443764.4582 0.5475 0.296 0.607 0.715
2443764.4623 0.5483 0.293 0.618 0.790
2443764.4651 0.5489 0.293 0.621 0.790
2443764.4693 0.5497 0.285 0.601 0.706
2443764.4720 0.5502 0.284 0.600 0.683
2443764.4755 0.5509 0.289 0.603 0.695
2443764.4790 0.5516 0.274 0.594 0.693
2443778.3439 0.2717 0.018 0.354 0.476
2443778.3466 0.2722 0.022 0.360 0.454
2443778.3501 0.2729 0.021 0.349 0.435
2443778.3529 0.2735 0.024 0.344 0.456
2443778.3557 0.2740 0.020 0.362 0.477
```

```
2443778.3591 0.2747 0.012 0.360 0.482
2443778.3619 0.2752 0.006 0.351 0.484
2443779.3029 0.4598 0.328 0.648 0.740
2443779.3057 0.4604 0.329 0.644 0.753
2443779.3091 0.4610 0.327 0.646 0.753
2443779.3119 0.4616 0.328 0.662 0.733
2443779.3147 0.4621 0.342 0.651 0.761
2443779.3182 0.4628 0.344 0.657 0.756
2443779.3210 0.4634 0.343 0.663 0.729
2443779.3237 0.4639 0.352 0.663 0.775
2443779.3272 0.4646 0.349 0.669 0.755
2443787.2884 0.0265 0.732 1.074 1.209
2443787.2912 0.0270 0.722 1.063 1.211
2443787.2940 0.0276 0.713 1.059 1.183
2443787.2974 0.0282 0.695 1.054 1.175
2443787.3009 0.0289 0.685 1.037 1.193
2443787.3044 0.0296 0.676 1.040 1.168
2443787.3079 0.0303 0.675 1.026 1.175
2443787.3113 0.0309 0.660 1.018 1.132
2443787.3131 0.0313 0.656 0.999 1.138
2443787.3176 0.0322 0.641 0.980   -
2443787.3242 0.0335 0.632 0.982 1.118
2443787.3280 0.0342 0.617 0.953 1.076
2443787.3329 0.0352 0.609 0.947 1.070
2443787.3349 0.0356 0.598 0.946 1.126
2443787.3412 0.0368 0.579 0.929 1.044
2443787.3440 0.0374 0.579 0.930 1.041
2443787.3481 0.0382 0.568 0.904 1.006
2443787.3530 0.0391 0.551 0.901 1.286
2443787.3558 0.0397 0.545 0.894 1.070
2443788.3281 0.2304 0.019 0.304 0.440
2443788.3316 0.2311 0.048 0.316 0.441
2443788.3350 0.2318 0.014 0.293 0.496
2443788.3378 0.2323 0.023 0.294 0.433
2443788.3413 0.2330 0.041 0.362   -
2443788.3455 0.2338 0.029 0.321 0.481
2443788.3496 0.2346 0.018 0.300   -
2443790.3809 0.6332 0.111 0.436 0.547
2443790.3837 0.6337 0.115 0.428 0.526
2443790.3871 0.6344 0.111 0.434 0.548
2443790.3899 0.6349 0.104 0.431 0.583
2443790.3927 0.6355 0.108 0.428 0.573
2443792.3976 0.0288 0.692 1.044 1.353
2443792.4004 0.0293 0.674 1.022 1.163
2443792.4039 0.0300 0.667 1.010 1.159
2443792.4067 0.0306 0.671 1.020 1.154
2443792.4101 0.0313 0.659 1.008 1.146
2443792.4136 0.0319 0.643 0.997 1.139
2443792.4164 0.0325 0.636 0.988 1.098
2443792.4219 0.0336 0.627 0.961 1.109
2443792.4247 0.0341 0.623 0.964 1.088
2443792.4275 0.0347 0.612 0.957 1.090
2443792.4310 0.0354 0.601 0.945 1.082
2443792.4338 0.0359 0.590 0.942 1.083
2443838.2146 0.0174 0.830 1.185 1.348
```

```
2443838.2174 0.0179  0.818  1.173  1.335
2443838.2202 0.0185  0.805  1.170  1.327
2443838.2236 0.0191  0.806  1.174  1.465
2443838.2271 0.0198  0.803  1.150  1.336
2443838.2299 0.0204  0.792  1.137  1.535
2443838.2327 0.0209  0.784  1.137  1.326
2443838.2354 0.0214  0.786  1.134  1.298
2443838.2382 0.0220  0.779  1.130  1.118
2443839.1952 0.2097  0.050  0.374  0.487
2443839.1979 0.2103  0.038  0.358  0.482
2443839.2007 0.2108  0.033  0.352  0.466
2443839.2035 0.2114  0.036  0.359  0.473
2443839.2063 0.2119  0.046  0.368  0.464
2443839.2090 0.2125  0.048  0.361  0.483
2443839.2132 0.2133  0.031  0.358    -
2443899.1630 0.9745  0.722  1.073  1.181
2443899.1658 0.9750  0.721  1.079  1.231
2443899.1699 0.9758  0.740  1.083  1.236
2443899.1734 0.9765  0.751  1.086  1.224
2443899.1762 0.9771  0.759  1.105  1.243
2443899.1831 0.9784  0.766  1.129  1.265
2443899.1866 0.9791  0.776  1.127  1.313
2443899.1915 0.9801  0.779  1.144  1.289
2443899.1936 0.9805  0.783  1.153  1.276
2443899.1970 0.9811  0.795  1.159  1.299
2443900.1602 0.1701  0.069  0.394  0.465
2443900.1630 0.1707  0.060  0.394  0.468
2443900.1658 0.1712  0.060  0.385  0.474
2443900.1699 0.1720  0.064  0.392  0.465
2443900.1727 0.1726  0.065  0.394  0.460
2443900.1762 0.1732  0.062  0.390  0.452
2443900.1790 0.1738  0.060  0.368  0.464
2443900.1817 0.1743  0.056  0.373  0.453
2443900.1845 0.1749  0.067  0.380  0.451
2443900.1873 0.1754  0.065  0.388  0.478
2443903.1697 0.7605  0.036  0.361  0.439
2443903.1725 0.7611  0.030  0.358  0.444
2443903.1760 0.7618  0.024  0.366  0.447
2443903.1802 0.7626  0.037  0.364  0.446
2443903.1899 0.7645    -    0.353  0.450
2443903.1927 0.7650  0.044  0.353  0.450
2443903.1961 0.7657  0.043  0.366  0.431
2443903.1989 0.7663  0.030  0.358  0.438
2444175.3407 0.1561  0.050  0.392  0.487
2444175.3442 0.1568  0.058  0.398  0.483
2444175.3504 0.1580  0.053  0.380  0.460
2444175.3532 0.1586  0.066  0.395  0.474
2444175.3567 0.1593  0.059  0.387  0.496
2444175.3601 0.1600  0.069  0.386    -
2444175.3629 0.1605  0.054  0.385    -
2444175.3664 0.1612  0.042    -      -
2444195.2223 0.0566  0.341  0.657  0.772
2444195.2251 0.0571  0.343  0.666  0.776
2444195.2286 0.0578  0.323  0.665  0.752
2444195.2314 0.0584  0.326  0.670  0.754
```

```
2444195.2348 0.0591  0.312  0.651  0.762
2444195.2383 0.0597  0.317  0.646  0.756
2444195.2446 0.0610  0.286  0.613  0.743
2444195.2473 0.0615  0.296  0.620  0.723
2444195.2508 0.0622  0.297  0.618  0.721
2444195.2536 0.0627  0.297  0.615  0.716
2444195.2564 0.0633  0.288  0.610  0.726
2444195.2592 0.0638  0.282  0.615  0.728
2444197.2890 0.4621  0.326  0.656  0.722
2444197.2918 0.4626  0.327  0.659  0.704
2444197.2946 0.4632  0.334  0.655  0.733
2444197.2973 0.4637  0.340  0.646  0.728
2444197.3008 0.4644  0.347  0.673  0.740
2444197.3064 0.4655  0.354  0.676  0.727
2444197.3091 0.4660  0.356  0.681  0.738
2444197.3119 0.4665  0.365  0.665  0.758
2444218.2205 0.5685  0.213  0.534  0.624
2444218.2239 0.5691  0.214  0.525  0.631
2444218.2267 0.5697  0.216  0.531  0.632
2444218.2295 0.5702  0.207  0.523  0.609
2444218.2323 0.5708  0.213  0.527  0.614
2444218.2350 0.5713  0.210  0.515  0.633
2444218.2392 0.5721  0.205  0.507  0.619
2444218.2420 0.5727  0.194  0.521  0.602
2444218.2448 0.5732  0.198  0.513  0.599
2444218.2475 0.5738  0.195  0.519  0.616
2444220.1696 0.9509  0.430  0.763  0.883
2444220.1723 0.9514  0.428  0.770  0.897
2444220.1758 0.9521  0.435  0.764  0.881
2444220.1786 0.9526  0.447  0.779  0.886
2444220.1814 0.9532  0.457  0.819  0.922
2444220.1848 0.9538  0.475  0.800  0.911
2444220.1876 0.9544  0.463  0.803  0.921
2444220.1904 0.9549  0.478    -    0.929
2444220.1932 0.9555  0.471  0.794  0.956
2444220.1959 0.9560  0.503  0.850  0.963
2444220.2001 0.9568  0.469  0.827  0.867
2444220.2036 0.9575  0.481  0.843  0.940
2444220.2064 0.9581  0.503  0.842  0.979
2444220.2091 0.9586  0.512  0.849  0.968
2444220.2119 0.9592  0.523  0.853  0.968
2444220.2598 0.9686  0.635  0.968  1.069
2444220.2626 0.9691  0.632  0.974  1.087
2444220.2661 0.9698  0.639  0.984  1.110
2444220.2696 0.9705  0.649  0.991  1.099
2444220.2723 0.9710  0.659  0.993  1.120
2444220.2758 0.9717  0.663  1.003  1.174
2444220.2786 0.9722  0.675  1.033  1.154
2444281.1864 0.9214  0.222  0.537  0.622
2444281.1899 0.9221  0.231  0.535  0.629
2444281.1933 0.9227  0.220  0.547  0.620
2444281.1961 0.9233  0.224  0.551  0.619
2444281.2010 0.9243  0.229  0.549  0.638
2444281.2038 0.9248  0.234  0.545  0.650
2444281.2072 0.9255  0.231  0.555  0.640
```

```
2444281.2135  0.9267  0.236  0.569  0.626
2444281.2170  0.9274  0.244  0.565  0.639
2444281.2197  0.9279  0.247  0.584  0.644
2444434.4183  0.9831  0.815  1.155  1.284
2444434.4217  0.9837  0.829  1.194  1.319
2444434.4245  0.9843  0.809  1.190  1.286
2444434.4301  0.9854  0.840  1.191  1.314
2444434.4349  0.9863  0.824  1.186  1.333
2444434.4377  0.9869  0.865  1.227  1.322
2444434.4405  0.9874  0.867  1.232  1.389
2444434.4446  0.9882  0.861  1.232  1.323
2444434.4481  0.9889  0.843  1.216  1.346
2444434.4509  0.9894  0.866  1.210  1.319
2444437.3899  0.5660  0.205  0.551  0.618
2444437.3944  0.5669  0.213  0.532  0.698
2444437.3968  0.5674  0.200  0.545  0.618
2444437.4128  0.5705  0.192  0.531  0.691
2444437.4163  0.5712  0.191  0.523  0.610
2444437.4197  0.5719  0.200  0.544  0.603
2444437.4225  0.5724  0.182  0.516  0.601
2444437.4253  0.5730  0.185  0.511  0.586
2444438.4039  0.7650  0.021  0.344  0.411
2444438.4066  0.7655  0.020  0.345  0.438
2444438.4094  0.7660  0.028  0.356  0.428
2444438.4122  0.7666  0.025  0.360  0.401
2444438.4150  0.7671  0.028  0.336  0.423
2444438.4178  0.7677  0.014  0.359  0.414
2444438.4205  0.7682  0.023  0.339  0.410
2444438.4233  0.7688  0.029  0.352  0.413
2444438.4261  0.7693  0.029  0.341  0.418
2444438.4289  0.7699  0.023  0.347  0.398
2444442.4279  0.5544  0.273  0.580  0.625
2444442.4307  0.5550  0.256  0.586  0.633
2444442.4341  0.5556  0.265  0.576  0.657
2444442.4369  0.5562  0.255  0.577  0.675
2444442.4397  0.5567  0.261  0.566  0.638
2444442.4445  0.5577  0.247  0.566  0.629
2444459.3408  0.8724  0.160  0.476  0.589
2444459.3443  0.8731  0.143  0.477  0.581
2444459.3477  0.8738  0.154  0.461  0.599
2444459.3505  0.8744  0.152  0.472  0.585
2444459.3540  0.8750  0.161  0.468  0.591
2444459.3568  0.8756  0.149  0.476  0.595
2444459.3595  0.8761  0.151  0.481  0.587
2444459.3686  0.8779  0.154  0.491  0.594
2444459.3713  0.8784  0.161  0.473  0.606
2444459.3748  0.8791  0.163  0.491  0.606
2444467.4043  0.4544  0.281  0.604  0.667
2444467.4077  0.4550  0.286  0.624  0.691
2444467.4112  0.4557  0.283  0.626  0.691
2444467.4140  0.4563  0.292  0.624  0.696
2444467.4168  0.4568  0.295  0.629  0.688
2444467.4202  0.4575  0.299  0.628  0.718
2444467.4230  0.4580  0.296  0.637  0.705
2444467.4265  0.4587  0.293  0.629  0.693
```

```
2444467.4300 0.4594 0.305 0.577 0.698
2444467.4327 0.4600 0.302 0.574 0.713
2444470.3913 0.0404 0.533 0.859 0.987
2444470.3941 0.0409 0.520 0.852 0.973
2444470.3975 0.0416 0.510 0.839 0.970
2444470.4003 0.0421 0.498 0.847 0.964
2444470.4038 0.0428 0.494 0.820 0.969
2444470.4073 0.0435 0.494 0.824 0.857
2444470.4107 0.0442 0.486 0.821 0.953
2444470.4135 0.0447 0.468 0.808 0.950
2444470.4170 0.0454 0.465 0.796 0.827
2444485.3272 0.9706 0.645 1.004 1.153
2444485.3321 0.9715 0.667 1.024 1.175
2444485.3369 0.9725 0.656 1.004 1.137
2444485.3404 0.9732 0.685 1.040 1.202
2444485.3439 0.9738 0.690 1.044 1.208
2444485.3473 0.9745 0.698 1.046 1.204
2444485.3508 0.9752 0.704 1.063 1.225
2444485.3543 0.9759 0.705 1.069 1.235
2444485.3571 0.9764 0.717 1.080 1.254
2444485.3598 0.9770 0.727 1.084 1.243
2444572.1798 0.0097 0.871 1.235 1.393
2444572.1860 0.0109 0.865 1.234 1.384
2444572.1888 0.0114 0.859 1.222 1.390
2444572.1916 0.0120 0.855 1.221 1.379
2444572.1944 0.0125 0.855 1.219 1.366
2444572.1971 0.0131 0.858 1.214 1.382
2444572.2048 0.0146 0.836 1.195 1.358
2444572.2159 0.0167 0.824 1.181 1.348
2444572.2194 0.0174 0.806 1.151 1.342
2444572.2221 0.0180 0.797 1.151 1.329
2444579.2172 0.3903 0.131 0.454 0.558
2444579.2200 0.3908 0.107 0.446 0.542
2444579.2213 0.3911 0.149 0.455 0.531
2444579.2234 0.3915 0.113 0.444 0.527
2444579.2276 0.3923 0.134 0.439 0.528
2444579.2299 0.3928 0.134 0.443 0.517
2444579.2318 0.3932 0.138 0.493 0.554
2444579.2338 0.3935 0.113 0.447 0.527
2444579.2359 0.3940   -   0.474 0.531
2444579.2389 0.3945   -   0.465 0.500
2444579.2415 0.3951   -     -   0.531
2444582.2294 0.9812 0.831 1.159 1.351
2444582.2329 0.9819 0.818 1.168 1.327
2444582.2391 0.9831 0.835 1.186 1.315
2444582.2433 0.9840 0.838 1.186 1.335
2444582.2461 0.9845 0.860 1.211 1.360
2444582.2495 0.9852 0.856 1.185 1.347
2444582.2530 0.9859 0.844 1.197 1.369
2444582.2565 0.9866 0.880 1.250 1.369
2444582.2600 0.9872 0.889 1.212 1.336
2444582.2634 0.9879 0.836 1.204 1.320
2444582.2662 0.9885 0.803   -     -
2444582.2697 0.9891 0.834 1.194 1.330
2444582.2759 0.9904 0.867   -   1.310
```

```
2444582.2794 0.9910 0.846 1.185 1.321
2444582.2822 0.9916 0.851 1.205 1.348
2444582.2891 0.9930 0.867   -     -
2444582.2933 0.9938 0.859 1.228 1.365
2444582.2961 0.9943 0.855 1.202 1.343
2444582.2995 0.9950 0.872 1.229 1.383
2444582.3023 0.9955 0.877 1.227 1.355
2444582.3065 0.9964 0.892 1.220 1.346
2444582.3099 0.9970 0.883 1.214 1.370
2444582.3127 0.9976 0.868 1.211 1.336
2444582.3169 0.9984 0.886 1.220 1.376
2444582.3204 0.9991 0.876 1.217 1.369
2444600.1684 0.5006 0.471 0.782 0.851
2444600.1718 0.5013 0.468 0.779 0.851
2444600.1746 0.5018 0.467 0.777 0.848
2444600.1774 0.5024 0.474 0.782 0.843
2444600.1802 0.5029 0.463 0.778 0.843
2444600.1843 0.5037 0.468 0.796 0.836
2444600.1871 0.5043 0.462 0.775 0.878
2444903.2114 0.9528 0.450 0.795 0.927
2444903.2149 0.9535 0.463 0.782 0.923
2444903.2177 0.9540 0.465 0.810 0.923
2444903.2212 0.9547 0.484 0.822 0.947
2444903.2281 0.9561 0.492 0.837 0.951
2444903.2316 0.9568 0.499 0.838 0.958
2444903.2343 0.9573 0.503 0.841 0.974
2444903.2378 0.9580 0.514 0.838 0.981
2444903.2427 0.9589 0.519 0.857 0.979
2444903.2462 0.9596 0.535 0.870 0.994
2444903.2489 0.9602 0.543 0.882 1.001
2444903.2524 0.9608 0.549 0.889 1.013
2444903.2594 0.9622 0.553 0.896 1.037
2444903.2628 0.9629 0.570 0.916 1.027
2444903.2670 0.9637 0.579 0.929 1.070
2444903.2698 0.9643 0.589 0.931 1.079
2444903.2767 0.9656 0.600 0.958 1.099
2444903.2802 0.9663 0.609 0.957 1.105
2444903.2837 0.9670 0.621 0.962 1.116
2444903.2864 0.9675 0.636 0.976 1.118
2444903.2885 0.9679 0.637 0.980 1.121
2444903.2913 0.9685 0.649 1.001 1.125
2444903.2948 0.9692 0.659 1.001 1.140
2444903.2968 0.9696 0.667 0.998 1.141
2444908.1747 0.9265 0.237 0.565 0.674
2444908.1782 0.9272 0.248 0.579 0.699
2444908.1817 0.9279 0.257 0.587 0.694
2444908.1838 0.9283 0.249 0.574 0.692
2444908.1872 0.9290 0.259 0.581 0.678
2444908.1900 0.9295 0.264 0.587 0.691
2444908.1935 0.9302 0.262 0.582 0.691
2444908.1969 0.9309 0.262 0.629 0.701
2444908.2011 0.9317 0.281 0.576 0.709
2444908.2046 0.9324 0.280 0.588 0.714
2444908.2115 0.9337 0.287 0.610 0.722
2444908.2150 0.9344 0.282 0.623 0.728
```

```
2444908.2178 0.9350  0.287  0.615  0.724
2444908.2206 0.9355  0.304  0.634  0.745
2444908.2240 0.9362  0.313  0.644  0.756
2444908.2310 0.9376  0.313  0.649  0.753
2444908.2338 0.9381  0.319  0.649  0.754
2444908.2365 0.9386  0.334  0.651  0.755
2444908.2393 0.9392  0.328  0.650  0.756
2444908.2421 0.9397  0.332  0.655  0.771
2444908.2456 0.9404  0.333  0.667  0.789
2444908.2497 0.9412  0.345  0.672  0.763
2444934.1513 0.0227  0.751  1.109  1.243
2444934.1541 0.0233  0.757  1.103  1.248
2444934.1569 0.0238  0.746  1.101  1.227
2444934.1603 0.0245  0.741  1.073  1.261
2444934.1687 0.0261  0.709  1.053  1.202
2444934.1721 0.0268  0.706  1.061  1.196
2444934.1749 0.0273  0.696  1.032  1.208
2444934.1777 0.0279  0.695  1.031  1.190
2444934.1812 0.0286  0.675  1.035  1.152
2444934.2346 0.0391  0.537  0.880  1.001
2444934.2381 0.0397  0.529  0.877  0.961
2444934.2409 0.0403  0.519  0.867  0.978
2444934.2437 0.0408  0.518  0.852  0.967
2444934.2471 0.0415  0.507  0.829  0.949
2444934.2499 0.0421  0.502  0.832  0.946
2444934.2534 0.0427  0.496  0.834  0.956
2444934.2555 0.0432  0.465  0.805  0.917
2444934.2590 0.0438  0.461  0.785  0.919
2444934.2617 0.0444  0.451  0.787  0.895
2444934.2645 0.0449  0.445  0.762  0.886
2444934.2680 0.0456  0.433  0.754  0.866
2444934.2742 0.0468  0.419  0.745  0.863
2444934.2770 0.0474  0.420  0.738  0.860
2444934.2805 0.0481  0.406  0.732  0.834
2444934.2846 0.0489  0.402  0.723  0.833
2444934.2874 0.0494  0.390  0.714  0.828
2444934.2909 0.0501  0.392  0.715  0.824
2444934.2937 0.0507  0.378  0.709  0.825
2444934.2944 0.0508  0.386  0.707  0.825
2444991.1805 0.2109  0.029  0.340  0.412
2444991.1853 0.2119  0.027  0.400  0.436
2444991.1881 0.2124  0.022  0.355  0.397
2444991.1909 0.2130  0.027  0.354  0.411
2444991.1943 0.2136  0.020  0.354  0.413
2444991.1971 0.2142  0.037  0.364  0.422
2444991.2027 0.2153  0.029  0.349  0.413
2444991.2062 0.2160  0.030  0.328  0.402
2444991.2089 0.2165  0.016  0.330  0.397
2444991.2124 0.2172  0.023  0.335  0.396
2444992.1714 0.4053  0.142  0.461  0.553
2444992.1742 0.4059  0.142  0.467  0.546
2444992.1771 0.4065  0.151  0.469  0.531
2444992.1798 0.4070  0.150  0.465  0.534
2444992.1732 0.4057  0.155  0.467  0.555
2444992.1860 0.4082  0.150  0.464  0.557
```

```
2444992.1895 0.4089 0.149 0.465 0.555
2444992.1923 0.4094 0.154 0.468 0.561
2444992.1957 0.4101 0.142 0.466 0.549
2444992.1992 0.4108 0.125 0.455 0.530
2445268.3768 0.5924 0.152 0.465 0.554
2445268.3796 0.5930 0.147 0.482 0.555
2445268.3831 0.5937 0.148 0.466 0.559
2445268.3858 0.5942 0.150 0.471 0.601
2445268.3873 0.5945 0.147 0.451 0.531
2445268.3921 0.5954 0.149 0.482 0.576
2445268.3942 0.5959 0.144 0.467 0.533
2445311.2428 0.0021 0.830   -   1.279
2445311.2476 0.0030 0.831 1.182 1.244
2445311.2525 0.0040 0.817 1.180 1.304
2445311.2574 0.0049 0.837 1.173 1.289
2445311.2622 0.0059 0.854 1.223 1.424
2445311.2650 0.0064 0.823 1.179 1.326
2445311.2699 0.0074 0.817 1.186 1.297
2445322.1395 0.1398 0.082 0.417 0.492
2445322.1423 0.1404 0.076 0.407 0.507
2445322.1451 0.1409 0.081 0.404 0.503
2445322.1478 0.1415 0.091 0.412 0.513
2445322.1520 0.1423 0.084 0.410 0.494
2445322.1541 0.1427 0.082 0.408 0.494
2445322.1569 0.1432 0.080 0.404 0.482
2445322.1596 0.1438 0.079 0.420 0.489
2445325.3220 0.7642 0.025 0.329 0.414
2445325.3255 0.7649 0.023 0.351 0.413
2445325.3290 0.7656 0.005 0.336 0.432
2445325.3318 0.7661 0.028 0.345 0.413
2445325.3352 0.7668 0.028 0.355 0.445
2445325.3380 0.7673 0.026 0.342 0.413
2445325.3415 0.7680 0.015 0.324 0.379
2445349.2841 0.4652 0.343 0.656 0.697
2445349.2869 0.4657 0.354 0.655 0.700
2445349.2904 0.4664 0.354 0.656 0.705
2445349.2939 0.4671 0.351 0.650 0.691
2445349.2973 0.4678 0.357 0.654 0.711
2445349.3008 0.4684 0.367 0.669 0.710
2445349.3036 0.4690 0.336 0.650 0.716
2445349.3070 0.4697 0.366 0.687 0.688
2445349.3098 0.4702 0.386 0.663 0.743
2445349.3140 0.4710 0.368 0.678 0.736
2445349.3168 0.4716 0.373 0.681 0.734
2445349.3195 0.4721 0.378 0.650 0.714
2445350.2834 0.6612 0.076 0.387 0.447
2445350.2868 0.6619 0.078 0.397 0.454
2445350.2911 0.6627 0.069 0.400 0.439
2445350.2939 0.6633 0.070 0.401 0.420
2445350.2980 0.6641 0.050 0.379 0.421
2445350.3008 0.6646 0.052 0.399 0.431
2445350.3036 0.6652 0.060 0.379 0.482
2445350.3064 0.6657 0.062 0.375 0.444
2445350.3098 0.6664 0.084 0.396 0.426
2445350.3140 0.6672 0.060 0.350 0.439
```

```
2445617.2093 0.0278  0.697  1.058  1.211
2445617.2121 0.0284  0.683  1.050  1.193
2445617.2156 0.0291  0.692  1.036  1.172
2445617.2184 0.0296  0.676  1.021  1.202
2445617.2218 0.0303  0.663  1.011  1.155
2445617.2246 0.0308  0.643  1.001  1.149
2445617.2274 0.0314  0.643  0.992  1.146
2445617.2309 0.0321  0.642  0.986  1.143
2446055.2053 0.9557  0.466  0.812  0.925
2446055.2090 0.9564  0.418  0.760  0.872
2446055.2170 0.9580  0.483  0.833  0.955
2446055.2199 0.9586  0.515  0.858  0.938
2446055.2225 0.9591  0.509  0.864  0.977
2446055.2255 0.9596  0.531  0.850  0.978
2446055.2287 0.9603  0.542  0.858  0.967
```